# Silicate glass fracture surface energy calculated from crystal structure and bond-energy data

M. Holzer [a], T. Waurischk [a], J. George [a,b], R. Maaß [a,c], R. Müller [a,*]

[a] Federal Institute of Materials Research and Testing (BAM), 12205 Berlin, Germany
[b] Institute of Condensed Matter Theory and Solid-State Optics at Friedrich Schiller University, 07743 Jena, Germany
[c] Department of Materials Science and Engineering, University of Illinois at Urbana-Champaign, Urbana, 61801 Illinois, USA

ABSTRACT

We present a novel method to predict the fracture surface energy, $\gamma$, of isochemically crystallizing silicate glasses using readily available crystallographic structure data of their crystalline counterpart and tabled diatomic chemical bond energies, $D^0$. The method assumes that $\gamma$ equals the fracture surface energy of the most likely cleavage plane of the crystal. Calculated values were in excellent agreement with those calculated from glass density, network connectivity and $D^0$ data in earlier work. This finding demonstrates a remarkable equivalence between crystal cleavage planes and glass fracture surfaces.

## 1. Introduction

The design of strong and tough oxide glasses remains a central goal in glass research for safer, more environmentally friendly, and thinner glass components [1,2]. Although several studies [3–8] provide important insights into composition - structure - property relations, this goal remains generally difficult to achieve due to the inherent brittleness of glasses at ambient temperature [9,10]. The search for promising candidates is further complicated since fracture toughness ($K_{Ic}$) measurements in brittle materials can be strongly affected by environmental conditions and other experimental factors [6,10–13].

Therefore, methods for predicting $K_{Ic}$ are of utmost importance for screening new promising candidates for tougher glasses. As a simple approach, $K_{Ic}$ of an ideal-brittle material can be estimated from fracture surface energy ($\gamma$), Young's modulus ($E$) and Poisson's ratio ($\nu$) in terms of the linear-elastic fracture mechanics for plane strain, according to [10]:

$$K_{Ic} = \sqrt{2\gamma E\left(1-\nu^2\right)^{-1}}. \quad (1)$$

Since the experimental $\gamma$ data required for such estimations are rare, models with predictive capabilities are crucial. One successful model was introduced by Rouxel [14] based on diatomic cation-oxide bond energy ($D_i^0$) and basic glass property data like molar mass ($M$) and density ($\rho$):

$$\gamma_\rho = \frac{1}{2}\left[\left(\frac{\rho}{M}N_A\right)^{2/3}\right]\cdot\sum_i \xi_i n_i D_i^0 N_A^{-1}, \quad (2)$$

where $N_A$ is Avogadro's constant and $\xi_i$ the atomic fraction of cation *i*. The term in brackets represents the number of bonds per fractured area, while the sum is their average molar bond energy, respectively. As a key aspect of the model, crack growth along the minimal possible energy path, i.e., along the weakest available bonds, is assumed. Thus, the switching parameter $n_i = \{0,1\}$ reflects whether or not the respective bond is involved in fracture. If the crack can proceed entirely breaking a certain low energy bond "*k*", $n_{i=k}$ is set to 1 and $n_{i\neq k}$ is set to 0. For alkaline and alkaline-earth silicate glasses, e.g., *k* could indicate $Na^+$-$O^-$ bonds and crack propagation entirely along these bonds is assumed possible if more than one non-bridging oxygen (*NBO*) occurs per one tetrahedral *[SiO4]* building block (*T*), (*NBO/T* > 1). In doing so, calculated $K_{Ic}$ results show excellent agreement with experimental data for more than 20 different glasses of known $E$ and $\rho$ values [14].

Alternative models utilize crystal structure data. For example, King et al. [15] used lattice constants and dissociation energies to compute $\gamma$ from bond per area data for brittle crystals and some amorphous materials. The model, however, tends to overestimate $\gamma$ for the glasses studied in Refs. [6,14]. Relying on the crystal-melt interface, Tielemann et al. [16] estimated $\gamma$ to predict the internal nucleation tendency for 20 isochemically crystallizing oxide glasses. Here, the crystal melt's interfacial energy was assumed to scale with the average crystal fracture surface energy of the low-indexed planes $\gamma_{001}$, $\gamma_{010}$, and $\gamma_{100}$ [16]. Similar to Ref. [15], crystal structure data were used to count the number density of ruptured bonds in these planes, and diatomic bond energies were used for $\gamma$ calculation. Adapting Rouxel's idea of crack growth along minimal energy paths [14], the faces parallel to (*001*), (*010*), and (*100*) with the lowest surface energy have been assumed to build the crystal surface.

* Corresponding author.
*E-mail address:* ralf.mueller@bam.de (R. Müller).

In the present work, we combine the approach of $\gamma$ calculation based on crystal structure data presented in Ref. [16] with the minimum energy crack path principle introduced in Ref. [14]. Thus, calculated $\gamma$-values were extracted from the lowest energy plane (cleavage plane) of an isochemical crystal, assuming crack propagation along this plane. Predicted $\gamma$-values do not only show excellent agreement with Ref. [14] but also unravel an interesting similarity between cleavage planes in a crystal and the fracture surface in their glassy state.

## 2. Theory

The present model assumes that the fracture surface energy of glasses, $\gamma$, is equal to that of the preferred cleavage plane of its isochemical crystalline counterpart, $\gamma_{hkl}$. This plane, (*hkl*), can be estimated in different ways. For some few cases, like sheet silicates with (*001*) cation layers, the cleavage plane can be easily anticipated to coincide with that plane [17]. Such a situation is exemplarily illustrated in Fig. 1 for Sanbornite.

Alternatively, (*hkl*) can be estimated from simple considerations within the unit cell as demonstrated in Ref. [16]. This procedure is based on counting the different cation-oxygen $M_i$-$O$ bonds (*i*) sliced by different planes within the unit cell and summing up the respective diatomic bond energy of monoxides, $D_i^0$, which were taken from Ref. [18]. This procedure, however, is very time-consuming and often challenging due to the required 3D situation awareness, in particular when many different or more complex structures are considered.

We therefore used the open-source software GALOCS [19] to predict the crystal's most likely cleavage plane (*hkl*). The software's geometrical algorithm lists the most promising cleavage candidates by locating lattice planes with the largest planar gaps (lengthwise) within a certain crystal structure. Then, VESTA 3 [20], which is also open-source, was used to identify the number and type of bonds involved during cleavage. The fracture surface energy, $\gamma_{hkl}$, based on this procedure was calculated as:

$$\gamma_{hkl} = \frac{1}{N_A}\sum_i \frac{s_{i,\ hkl}}{A_{hkl}} U_i, \tag{3}$$

where $A_{hkl}$ is the area of the (*hkl*) plane within the unit cell, $s_{i,hkl}$ is the number of broken cation-oxygen bonds $M_i$-$O$ in $A_{hkl}$, and $U_i$ is their bond energy calculated from $D_i^0$ according to:

$$U_i = \frac{yi/xi}{CN_i} D_i^0. \tag{4}$$

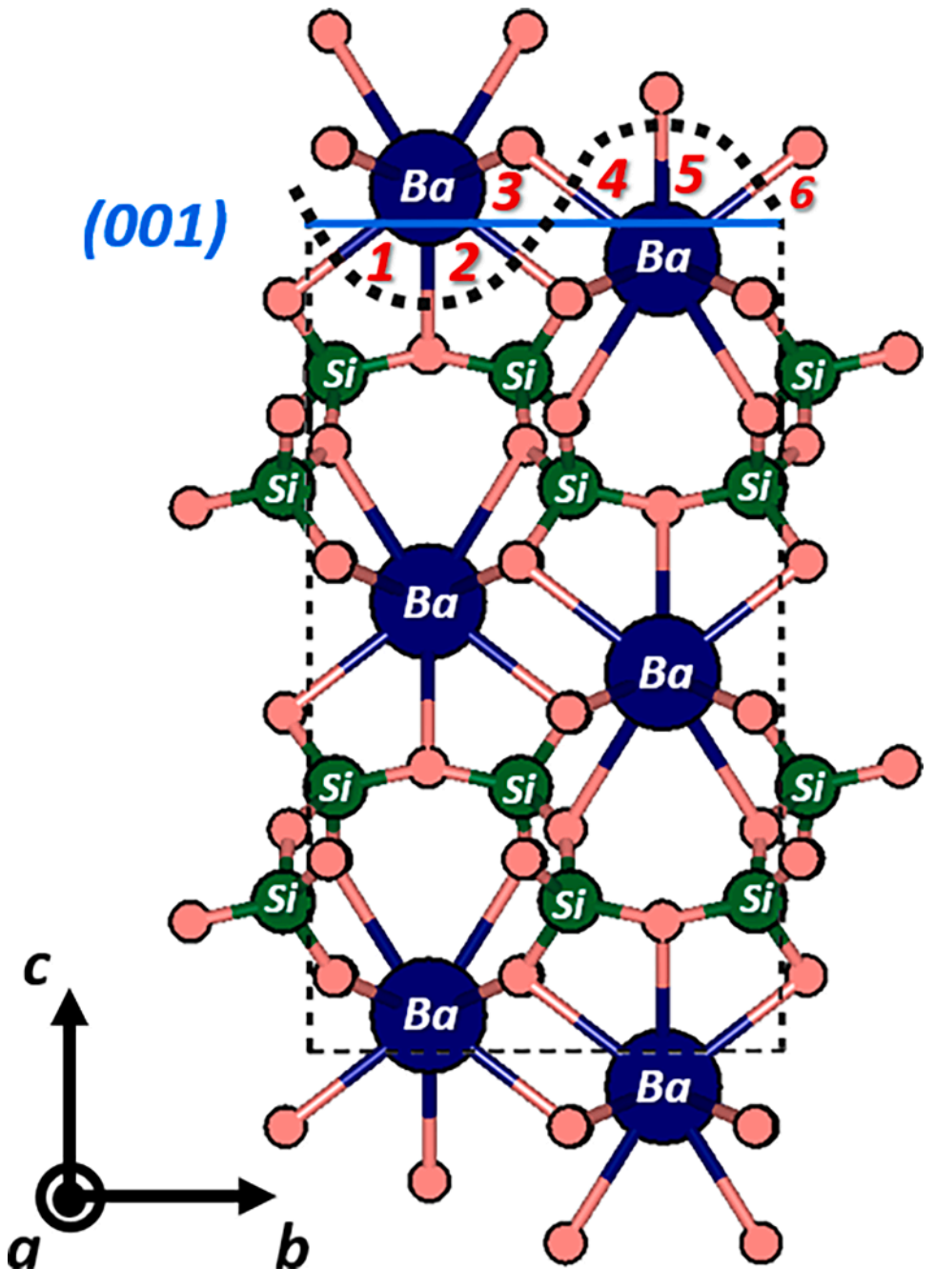

**Fig. 1.** VESTA 3 [20] illustration of the unit cell (dashed lines) of $BaS_2$ with the most likely cleavage plane (001) drawn in light blue. Six Ba-O bonds (red numbers) are cut during fracture. The cutting path is indicated by the dashed blue curve.

Here, $CN_i$ and $y_i/x_i$ represent the cation coordination number and the stoichiometric oxygen-to-metal-cation ratio available for $M_i$ bonding from the glass constituting oxide, $M_{ix}O_y$, respectively. Cations in different coordination environments are weighted according to their structure site fractions. The calculated values of $U_i$ are listed in Table 1. According to [16], Eqs. (3) and (4) assume an equally distributed bond energy within the coordination polyhedron. It furthermore entails that the bond energy of a given $M_i$-$O$ bond within the crystal structure is given by the respective diatomic one, corrected by the stoichiometric oxygen-to-metal-cation ratio available for $M_i$ bonding.

## 3. Calculations

### 3.1. Fracture surface energy calculated from crystal cleavage data, $\gamma_{hkl}$

Using the concept introduced above, the glass fracture surface energy, $\gamma$, was approximated as that of the most likely cleavage plane in its isochemical crystal, $\gamma_{hkl}$. As a first step, (*hkl*) were determined by GALOCS's processing of crystal lattice data (CIF files, see supplementary). Next, all (*hkl*)- specified cut bonds within the unit cell were added up according to Eqs. (3) and (4). In case of several cleavage planes, the one with the minimum $\gamma$- value was considered.

Fig. 1 illustrates this procedure for Sanbornite ($BaO{\cdot}2SiO_2$). GALOCS has identified (*001*) as the most probable cleavage plane (light blue line). The bold dashed curve shows the 6 Ba-O bonds cut by a fracture along (*001*) within the unit cell, which gives $s_{Ba,(001)} = 6$ in Eq. (3). As Ba has an oxygen coordination number of $CN_{Ba} = 9$, $D_{Ba}^0 = 561.9$ kJ mol$^{-1}$, and $y/x = 1$, Eq. (4) yields $U_{Ba} = 62.4$ kJ mol$^{-1}$. With this and $A_{(001)} = 0.36$ nm$^2$, Eq. (3) returns $\gamma_{hkl} = 1.74$ J m$^{-2}$.

This procedure was repeated for all 26 isochemical systems listed in Table 2. Using the data contained in Table 2, $\gamma_{hkl}$- values ranging between 0.59 J m$^{-2}$ (Sodium disilicate) and 5.18 J m$^{-2}$ (Boron oxide) were obtained.

**Table 1**
Molar diatomic bond energies, $D_i^0$, preferred oxygen coordination numbers, $CN_i$, oxygen-to-cation ratios ($y_i/x_i$) of constituting oxides of the isochemical glasses $M_{ix}O_y$, and bond energies, $U_i$, as calculated with Eq. (4).

| $M_i$ | $D_i^0$ (kJ mol$^{-1}$) | $CN_i$ | $y_i/x_i$ | $U_i$ (kJ mol$^{-1}$) |
|---|---|---|---|---|
| B | 808.9 | 3 | 1.5 | 404.5 |
| Si | 799.6 | 4 | 2.0 | 399.8 |
| Ti | 672.4 | 6 | 2.0 | 224.1 |
| Ba | 561.9 | 8 | 1.0 | 70.2 |
| Al | 511.0 | 4 | 1.5 | 191.6 |
| Ca | 402.1 | 8 | 1.0 | 50.3 |
| Pb | 382.0 | 6 | 1.0 | 63.7 |
| Mg | 363.2 | 6 | 1.0 | 60.5 |
| Li | 333.5 | 4 | 0.5 | 41.7 |
| Na | 256.1 | 6 | 0.5 | 21.3 |

**Table 2**
Glass and isochemical crystal data used for fracture surface energy calculation. Data references are listed in the Supplementary. $\rho$ = glass density, $\gamma_\rho$ = fracture surface energy calculated according to Eq. (2), (*hkl*) = preferred cleavage plane obtained from GALOCS, $CN_i$ = coordination number of cation *i*, $\gamma_{hkl}$ = fracture surface energy calculated according to Eqs. (3) and (4), *E* = Youngś modulus, $\nu$ = Poissonś ratio.

| Crystal | | $\rho$ (g cm$^{-3}$) | $\gamma_\rho$ (J m$^{-2}$) | (hkl) | *CN* | $\gamma_{hkl}$ (J m$^{-2}$) | *E* (GPa) | $\nu$ |
|---|---|---|---|---|---|---|---|---|
| $A_3S_2$ | Mullite | 2.76 | 3.44 | $(\bar{1}00)$ | 4–6/4 | 3.90 | 85 | 0.17 |
| B | Boron oxide | 1.85 | 4.99 | $(10\bar{2})$ | 3 | 5.18 | 17 | 0.26 |
| BaAS | Hexacelsian | 3.39 | 3.46 | (001) | 6/4/4 | 3.84 | 83 | 0.17 |
| $BaM_2A_3S_9$ | Barium osumilite | 2.76 | 3.50 | (010),(100) | 12/6/4/4 | 3.52 | 83 | 0.17 |
| $BaS_2$ | Sanbornite | 3,72 | 0.95 | (001) | 9/4 | 1.74 | 63 | 0.25 |
| $Ba_5S_8$ | Barium silicate | 3.93 | 3.62 | (001) | 6–8/4 | 3.17 | 72 | 0.26 |
| $BaTS_2$ | Fresnoite | 4.05 | 1.92 | (001) | 8/5/4 | 1.93 | 101 | 0.21 |
| $CAS_6$ | Anorthite | 2.7 | 3.44 | (001) | 6–7/4/4 | 3.14 | 94 | 0.17 |
| CS | Wollastonite | 2.9 | 1.19 | (001) | 8/4 | 1.12 | 93 | 0.20 |
| $LAS_2$ | β-Eucryptite | 2.43 | 3.64 | $(\bar{1}22)$ | 4/4/4 | 4.22 | 83 | 0.17 |
| $LAS_4$ | β-Spodumene | 2.37 | 3.65 | (100) | 6/5/4 | 3.45 | 86 | 0.17 |
| LS | Lithium metasilicate | 2.37 | 1.92 | (010) | 5/4 | 1.77 | 83 | 0.24 |
| $LS_2$ | Lithium disilicate | 2.35 | 1.19 | (010) | 4/4 | 0.99 | 79 | 0.22 |
| MA | Spinel | 2.7 | 3.00 | (100),(010),(001) | 4/6 | 2.71 | 83 | 0.17 |
| $M_2A_2S_5$ | Cordierite | 2.42 | 3.53 | (100) | 6/4/4 | 3.96 | 97 | 0.17 |
| $MCS_2$ | Diopside | 2.85 | 1.17 | $(1\bar{1}0)$ | 6/8/4 | 1.06 | 82 | 0.27 |
| MS | Enstatite | 2.58 | 1.10 | (110) | 6/4 | 1.31 | 107 | 0.27 |
| $M_2S$ | Forsterite | 2.65 | 1.59 | $(10\bar{1})$ | 6/4 | 2.070 | 128 | 0.25 |
| $NAS_2$ | Nepheline | 2.46 | 3.49 | (001) | 4/4/4 | 4.01 | 72 | 0.19 |
| $NAS_4$ | Jadeite | 2.53 | 3.39 | (001) | 8/4/4 | 3.38 | 71 | 0.19 |
| $NAS_6$ | Albite | 2.28 | 3.36 | (001) | 5/4/4 | 3.16 | 97 | 0.18 |
| NS | Sodium metasilicate | 2.56 | 1.27 | (010) | 5/4 | 1.47 | 54 | 0.22 |
| $NS_2$ | Sodium disilicate | 2.49 | 0.83 | (100) | 5–6/4 | 0.59 | 58 | 0.24 |
| PS | Alamosite | 5.98 | 3.13 | $(10\bar{1})$ | 4–5 | 3.26 | 45 | 0.27 |
| S | High quarz ($S_{HQ}$) | 2.2 | 3.62 | $(10\bar{1})$ | 4 | 3.60 | 70 | 0.15 |
| S | High cristobalite ($S_{HC}$) | 2.2 | 3.62 | $(10\bar{1})$ | 4 | 3.59 | 72 | 0.15 |

Short oxide notation used in Table 2, Figs. 2 and 3:
$A_3S_2 = 3Al_2O_3{\cdot}2SiO_2$, $B = B_2O_3$, $BaAS = BaO{\cdot}Al_2O_3{\cdot}SiO_2$, $BaM_2A_3S_9 = BaO{\cdot}2MgO{\cdot}3Al_2O_3{\cdot}9SiO_2$,
$BaS_2 = BaO{\cdot}2SiO_2$, $Ba_5S_8 = 5BaO{\cdot}8SiO_2$, $BaTS_2 = BaO{\cdot}TiO_2{\cdot}2SiO_2$, $CAS_6 = CaO{\cdot}Al_2O_3.6SiO_2$,
$CS = CaO{\cdot}SiO_2$, $LAS_2 = LiO{\cdot}Al_2O_3{\cdot}2SiO_2$, $LAS_4 = LiO{\cdot}Al_2O_3{\cdot}4SiO_2$, $LS = LiO{\cdot}SiO_2$, $LS_2 = LiO{\cdot}2SiO_2$,
$MA = MgO{\cdot}Al_2O_3$, $M_2A_2S_5 = 2MgO{\cdot}2Al_2O_3{\cdot}5SiO_2$, $MCS_2 = MgO{\cdot}CaO{\cdot}2SiO_2$, $MS = MgO{\cdot}SiO_2$,
$M_2S = 2MgO{\cdot}SiO_2$, $NAS_2 = Na_2O{\cdot}Al_2O_3{\cdot}2SiO_2$, $NAS_4 = Na_2O{\cdot}Al_2O_3{\cdot}4SiO_2$,
$NAS_6 = Na_2O{\cdot}Al_2O_3{\cdot}6SiO_2$, $NS = Na_2O{\cdot}SiO_2$, $NS_2 = Na_2O{\cdot}2SiO_2$, $PS = PbO{\cdot}SiO_2$, $S = SiO_2$.

### 3.2. Fracture surface energy calculated from glass properties, $\gamma_\rho$

Since experimental data for the surface energy of silicate glasses in which isochemical crystallization is observed are rare, a validation of the above approach was done by comparing $\gamma_{hkl}$ with $\gamma_\rho$ calculated from $\rho$, *NBO/T*, and $D_i^0$ according to Ref. [14]. This relies on the fact that $\gamma_\rho$ calculated via Eq. (2) provides a good approximation for the measured fracture surface energies $\gamma$ of several glasses, especially silicates [14]. To this end, we calculate $\gamma_\rho$ using the literature data summarized in Table 2 for 26 glass-forming systems. Table 2 further lists the Youngs moduli, *E*, and Poisson's ratio, $\nu$, for a later discussion on the fracture toughness of the different glasses. The good agreement between $\gamma_\rho$ and $\gamma_{hkl}$ becomes apparent in Fig. 2 which displays a tight correlation of both quantities.

For fused silica, both values also agree with the experimental fracture surface energy value for fused silica, $\gamma_{exp} = 3.48$ J m$^{-2}$ [21], see pentagon in Fig. 2. This is the only applicable experimental value to the authors best knowledge. In this case, $\gamma_{hkl} = 3.59$ J m$^{-2}$ was found for high cristobalite and 3.6 J m$^{-2}$ for high quartz whereas $\gamma_\rho = 3.62$ J m$^{-2}$ was obtained from Eq. (2) for fused silica.

## 4. Discussion

Whilst the data in Fig. 2 clearly demonstrates an equivalence between the fracture surface energy derived from the glass density and network connectivity data and the corresponding cleavage-plane fracture surface energy of the glass's isochemical crystalline phase, several consequences emerge.

Firstly, Fig. 2 demonstrates a separation into two groups of data. Half-filled symbols (< 2 J m$^{-2}$) correspond to glasses not fully polymerized ($NBO/T \geq 1$), while filled symbols (> 3 J m$^{-2}$) represent highly polymerized glasses ($NBO/T < 1$). This finding is intuitive since *NBO* indicates weaker-bonded non-bridging oxygen (as in *Na-O* with $U_i$ = 21.3 kJ mol$^{-1}$, Table 1), whereas *T* indicates the stronger-bonded tetrahedral building blocks of the silicate structure, which are mutually linked via bridging oxygens (as in *Si-O* with $U_i$ = 404.5 kJ mol$^{-1}$, Table 1).

Beyond this intuitive trend, however, it is remarkable to find a gap between both groups of data. Such anomalous jump of toughness-related properties was also reported in other studies and can be explained in terms of a percolation threshold for crack propagation along weaker bonds [5,22,23] according to the Topological Constraint Theory [24, 25]. While in Ref. [14] the percolation threshold is set to $NBO/T = 1$, we propose $NBO/T = 0.5$ as an alternative percolation threshold. According to the Topological Constraint Theory, this threshold corresponds to an average atomic coordination number $<r> = 2.4$ (Appendix A). Above this value, the crack can percolate via *NBO* bonds through the glass network. Although this little shift of the assumed percolation threshold affects only mullite in Fig. 2, the upper right position of mullite glass ($A_3S_2$) with $NBO/T = 0.5$ [26] might suggest that this criterion may be better suited in the present context.

Secondly, and most remarkable, $\gamma_\rho \approx \gamma_{hkl}$ indicates an equivalence, unexpected at first sight, between crystal cleavage planes and glass fracture surfaces. Indeed, it emphasizes the general applicability of considering crack propagation along the weakest bonding environments. Given the low sensitivity of $\gamma$ to a crystalline or glassy structural arrangement of the same bonds and a similar density as mostly given [27,28], $\gamma_\rho \approx \gamma_{hkl}$ suggests that the averaging applied to the glassy phase effectively reflects a chemically similar cleavage plane in the crystal [29]. Also, the percolation threshold for crack propagation along the weakest bonds ($NBO/T > 0.5$) seems to allow forming a cleavage plane of similar chemistry in the crystalline case.

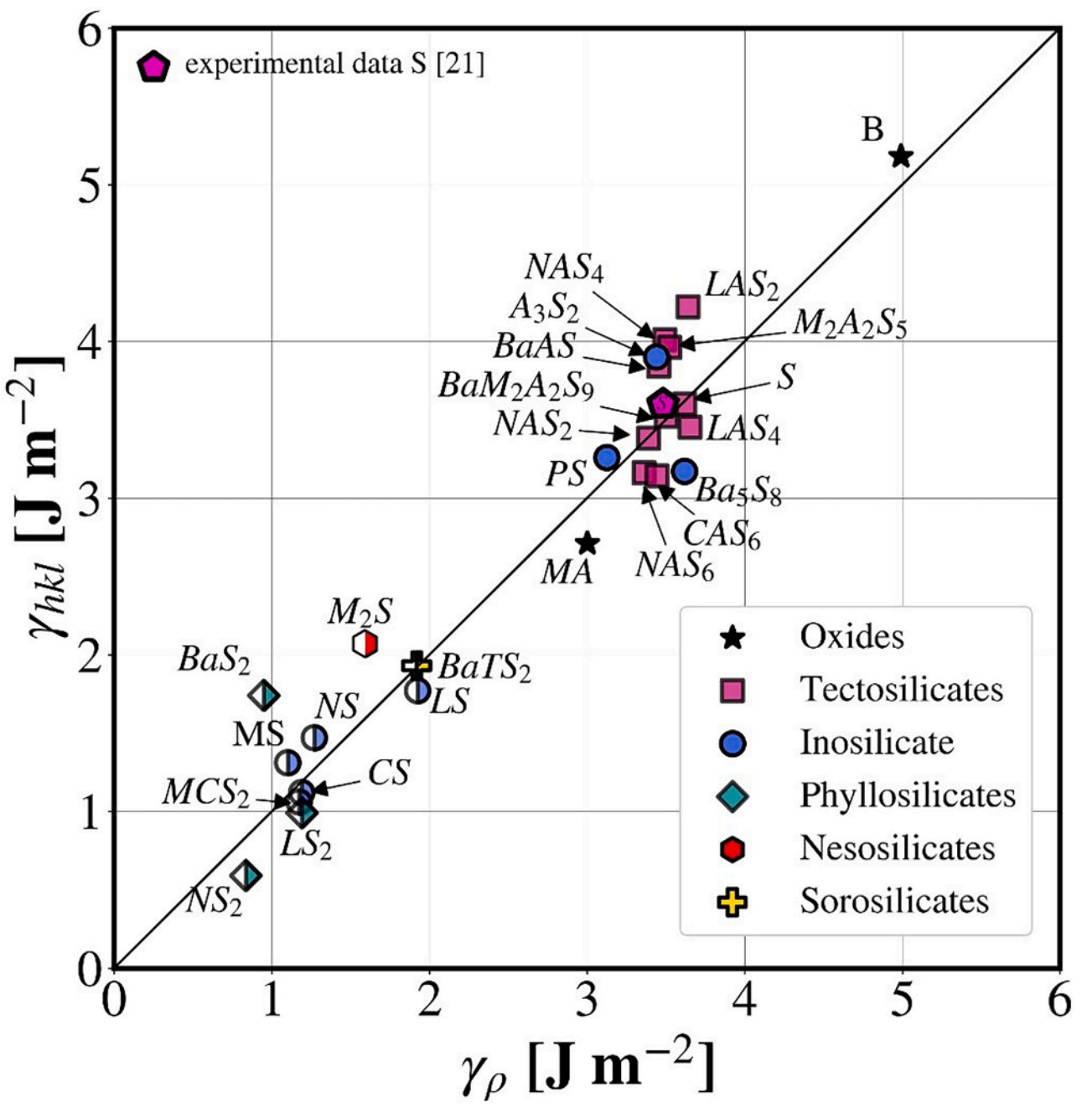

**Fig. 2.** Correlation between the presented fracture surface energy calculation methods, i.e., via crystal structure data, $\gamma_{hkl}$ Eqs. (3) and (4), versus values obtained from glass data, $\gamma_\rho$ (Eq. (2)). Data are listed in Table 2. Filled and half-filled symbols indicate highly ($NBO/T < 1$) and lowly polymerized structures. ($NBO/T \geq 1$), respectively. Silicate structure types are marked with different symbols and colors, whereas oxide structures are shown as black stars. The pentagon represent the only available experimental value, fused silica [21].

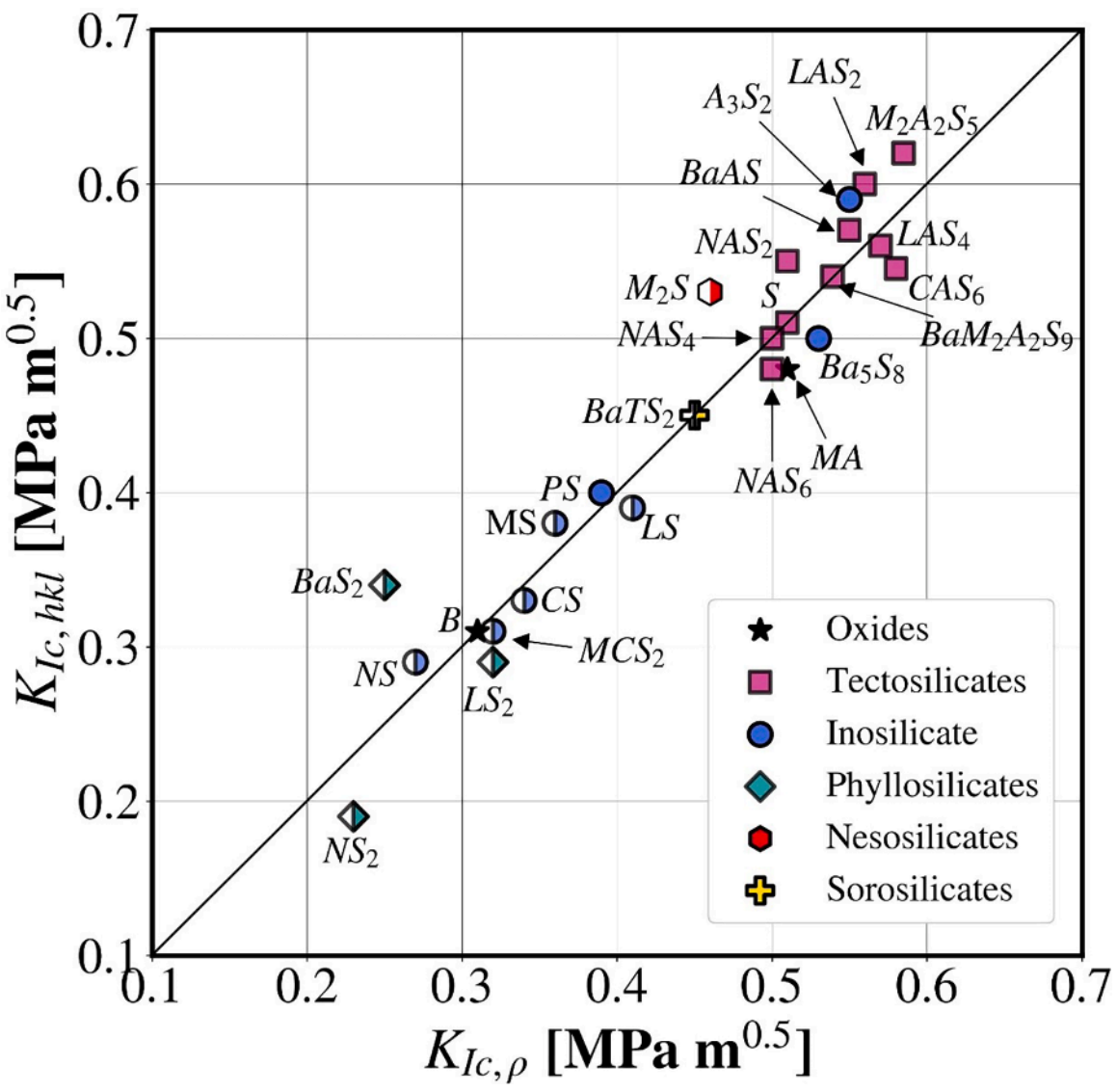

**Fig. 3.** Predicted $K_{Ic,hkl}$ values for the isochemical crystal obtained from $\gamma_{hkl}$ versus $K_{Ic,\rho}$ obtained from $\gamma_\rho$. See Table 2 caption for point labels and data sources.

This naturally leads to the question of whether $\gamma$ is systematically affected by the 6 silicate crystal structure types defined according to the different degrees of polymerization of their basic tetrahedral $SiO_4$ and $AlO_4$ building blocks [30]. These structure types are marked with different symbol shapes and colors in Fig. 2. As intuitively expected, *phyllo-*, *neso-* and *sorosilicates* (diamonds, hexagons, crosses) can be found in the lower range of $\gamma_{hkl}$ and $\gamma_\rho$. For those silicates, easy crack percolation is supported by weakly occupied layers (*phyllosilicates*) and/or the high amount of weaker bonds between isolated (*nesosilicates*) or paired building blocks (*sorosilicates*). *Tecto-* (or framework) *silicates* (squares) and their fully polymerized glassy counterparts, on the other hand, occur in the upper range of $\gamma_{hkl}$ and $\gamma_\rho$. In this case, easy crack growth along particularly weak chemical bonds (e.g., $Na\text{–}O^-$) is unlikely. Since their rings are cross- linked via $SiO_4^-$ and $AlO_4^-$ building blocks, cordierite ($M_2A_2S_5$) and Ba-Osumilite ($BaM_2A_2S_9$) can be treated as tectosilicates [30]. *Inosilicates* (circles) spread over the entire range of $\gamma_{hkl}$ and $\gamma_\rho$. Whereas single- chain silicates (pyroxenes) occur in the lower $\gamma$ region, multiple-chain silicates (PS, $B_5S_8$ and $A_3S_2$), with thus higher polymerization, can be found in the upper region.

Thirdly, using the derived and validated data for $\gamma_{hkl}$, it is now possible to use Eq. (1) for an evaluation of $K_{Ic}$. Fig. 3 shows $K_{Ic}$ values determined for the 26 glasses of Table 2 according to Eq. (1), using respective $E$ data, also listed in this table. $K_{Ic,hkl}$ was calculated from $\gamma_\rho$ according to Eq. (2) (abscissa). Despite a few exceptions (B, $M_2S$), comparing Fig. 2 with Fig. 3 illustrates the dominating effect of $\gamma$ on $K_{Ic}$. This finding reflects that $\gamma \approx \gamma_\rho \approx \gamma_{hkl}$ ranges between 0.59 and 5.18 J m$^{-2}$ (factor 7), whereas $E$ ranges between 45 and 128 GPa (factor 3, Table 2). The 25 $K_{Ic}$ values calculated from $\gamma_{hkl}$ and $\gamma_\rho$ both range between 0.2 to 0.6 MPa m$^{-1/2}$, which is very similar to that observed for the 20 glasses considered in Ref. [14], where experimental and predicted $K_{Ic}$ values range from 0.33 to 0.79 MPa m$^{-1/2}$. Since $\gamma_\rho$ allows for stronger manipulation of $K_{Ic}$ than $E$ according to Eq. (1), it becomes apparent that substantial toughness improvements are not expected by simple chemical substitution and its effect on singular bond energies for most silicate glasses. Instead, structural mechanisms, as discussed in [7, 8,31–34], must be considered, in which dissipation beyond bond rupture at the immediate crack tip becomes a contributing factor to internal stress relaxation and, therefore, $K_{Ic}$.

## 5. Conclusion

This work presents an easy-to-apply method to predict the glass fracture surface energy, $\gamma$, of ionocovalent isochemically crystallizing glasses from diatomic bond energies, $D_i^0$, crystal structure data, and the most likely crystal cleavage plane *(hkl)* computationally predicted by GALOCS [19].

In doing so, calculated data, $\gamma_{hkl}$, were found to excellently agree with surface energy data derived according to the approach suggested in Ref. [14], where $D_i^0$, network connectivity data and experimental glass density data, $\rho$, is used. The latter approach is known to successfully predict the fracture toughness, $K_{Ic}$, of oxide glasses. Demonstrating here that $\gamma \approx \gamma_\rho \approx \gamma_{hkl}$, indicates a remarkable equivalence between crystal cleavage planes and glass fracture surfaces. It also emphasizes the general validity of crack propagation along the weakest possible bonds available, irrespective of a glassy or crystalline structure.

Our proposed method broadens the available toolboxes for fracture toughness predictions of tough glass candidates, for which density or network connectivity data are not known and/or hard to predict. The here newly developed approach and earlier known methods have in common that they indicate a limited range of tuning $K_{Ic}$ across many possible bonding environments. As such, overall fracture toughness enhancements are likely only in reach if other structural dissipation mechanisms are exploited. Being currently limited to the case of isochemically crystallizing glass systems, the next steps naturally include an extension to non-isochemically crystallizing glasses. This may be achievable by combining our approach with principles from Conradt's constitutional crystal phase theory [35,36], eventually allowing predictions for any given glass composition.

## 6. Notations

| Symbol | Meaning | Unit |
|---|---|---|
| $\gamma$ | Fracture surface energy of an oxide glass | J $m^{-2}$ |
| $\gamma_\rho$ | $\gamma$ calculated via glass properties [14] | J $m^{-2}$ |
| $\gamma_{hkl}$ | $\gamma$ obtained from cleavage planes of its isochemical crystal | J $m^{-2}$ |
| $\xi_i$ | Atomic fraction of cation i | |
| $\nu$ | Poisson's ratio | |
| $\rho$ | Density | [kg $m^{-3}$] |
| $A_{hkl}$ | Projected area of *(hkl)* plane within the unit cell | [$m^2$] |
| $CN_i$ | Coordination number of cation i in a broken cation-oxygen bond $(M_i - O)$ | |
| $D_i^0$ | Diatomic bond energy of a monoxide | [kJ $mol^{-1}$] |
| $E$ | Young's Modulus | [GPa] |
| *(hkl)* | Miller indices of a crystal plane | |
| $K_{Ic}$ | Fracture toughness of an oxide glass | [MPa $m^{-0.5}$] |
| $K_{Ic,\ \rho}$ | $K_{Ic}$ – values calculated with $\gamma_\rho$ | [MPa $m^{-0.5}$] |
| $K_{Ic,\ hkl}$ | $K_{Ic}$ – values calculated with $\gamma_{hkl}$ | [MPa $m^{-0.5}$] |
| $M$ | Molar mass | [kg $mol^{-1}$] |
| $M_i - O$ | Cation – Oxygen bond | |
| $M_{ix}O_y$ | Constituting oxide | |
| $N_A$ | Avogadro constant | [$mol^{-1}$] |
| $n_i$ | Switching parameter reflecting if respective bond of cation i is involved in fracture | |
| $s_{i,hkl}$ | Number of broken cation-oxygen bonds $M_i$-$O$ in $A_{hkl}$ | |
| $U_i$ | Bond energy of broken cation-oxygen bonds $M_i$-$O$ in $A_{hkl}$ | [kJ $mol^{-1}$] |
| $y_i/x_i$ | stoichiometric oxygen-to-metal-cation ratio available for $M_i$ bonding from $M_{ix}O_y$ | |

(continued on next column)

(continued)

## Declaration of Competing Interest

The authors declare that they have no known competing financial interests or personal relationships that could have appeared to influence the work reported in this paper.

## Data availability

Data will be made available on request.

## Acknowledgments

This work was supported by the Federal Institute of Materials Research and Testing (BAM), "Förderprogramm Menschen – Ideen". We also want to thank Sydney Corona for proofreading and fruitful discussions.

## Supplementary materials

Supplementary material associated with this article can be found, in the online version, at doi:10.1016/j.jnoncrysol.2023.122679.

## Appendix A

According to [A1], the atomic structure of glass networks is rigid when the average number of constraints per atom, $n$, surpasses its degree of freedom. Dependent on $n$, such networks are either classified as flexible, isostatic, or stressed-rigid when $n < 3$, $n = 3$ or $n > 3$, respectively. Stress-rigid indicates that the structure is over-constrained resulting in internal stresses, while flexible networks maintain insufficient constraints for stabilization. At $n = 3$, the case of an isostatic system, the network is optimal (e.g., glass-forming ability).

Atomic constraints typically include bond-stretching (*BS*) and bond-bending (*BB*). In a fully polymerized covalent structure, where *BS* and *BB* constraints both apply for any atom, $n$ can be conveniently calculated from the average atomic coordination number $\langle r \rangle$, according to Maxwell's stability criterion for mechanical trusses [A1]:

$$n = \frac{5}{2}\langle r \rangle - 3 \tag{A-1}$$

For isostatic networks ($n = 3$), we obtain an $\langle r \rangle$ value of 2.4 which is called the rigidity percolation threshold. In this case, a rigid structure will percolate through the entire glass [A2], leading to an ideal glass network [A1, A3]. For structures involving different types or structural sites of atoms differently constrained, Eq. (A-1) is, however, not applicable. Instead, $n_i^{BB}$ and $n_i^{BS}$ must be derived individually and summarized weighed by their atomic fraction $\xi_i$ [A4]

$$n = \sum_i n_i\ \xi_i \tag{A-2}$$

Table A1 illustrates this procedure for binary alkali silicate glasses ($M_{[x]}$ - $SiO_{2[1-x]}$), where $X_i$ is the molar fraction of the species *i*. $\xi_i$ is calculated by dividing $X_i$ by the total mol number of atoms $N$. $\langle r_i \rangle$ gives the respective average species coordination number. $n^{BS}$ and $n^{BB}$ are then calculated from $\langle r_i \rangle$ according to the rules for $n_i^{BS}$ and $n_i^{BB}$ in [A4]. For the ionic bonds between $M$ and $NBO$, however, $n_i^{BB}$ was set to zero.

**Table A1**
Parameters for n calculation according to Eqs. (A-1) and (A-2) for a binary alkali silicate glass.

| $i$ | $X_i$ | $\langle r_i \rangle$ | $n_i^{BS}$ | $n_i^{BB}$ | $n_i$ |
|---|---|---|---|---|---|
| Si | (1 – x) | 4 | 2 | 5 | 7 |
| BO | (2 – 3x) | 2 | 1 | 1 | 2 |
| M | 2x | 1 | 0.5 | 0 | 0.5 |
| NBO | 2x | 2 | 1 | 0 | 1 |

*BO … bridging oxygen, NBO … non-bridging oxygen, M … alkali atom*

With $X_i$ and $n_i$ listed in Table A1, Eq. (A-2) yields:

$$n(x) = \frac{7(1-x) + 2(2-3x) + 0.5\ (2x) + 2x}{(1-x) + 2x + (2-x)} = \frac{(11-10x)}{3} \tag{A-3}$$

Solving Eq. (A-3) for $n = 3$, we obtain $x = 0.2$, which shows that for $x < 20$ mol% $M_2O$ a rigid glass structure is expected [A4]. This means that a crack percolation path formation is not possible. Finally, we determine the ratio of non-bridging oxygen atoms per *SiO₄* tetrahedra (*NBO/T*) for this composition, according to [A5]:

$$\frac{NBO}{T} = 2\frac{x_{Na2O}}{x_{SiO2}} \tag{A-4}$$

Thus, $NBO/T = 0.5$ indicates the structure rigidity percolation threshold of described binary system, which may also represent the percolation threshold for crack propagation along the weakest bonds without breaking the more energetically expensive *Si-O* bonds.